\documentclass[lettersize,journal]{IEEEtran}
\usepackage{amsmath,amsfonts}
\usepackage{algorithmic}
\usepackage{algorithm}
\usepackage{array}
\usepackage[caption=false,font=normalsize,labelfont=sf,textfont=sf]{subfig}
\usepackage{url}
\usepackage{verbatim}
\usepackage{graphicx}
\usepackage{cite}
\usepackage{xcolor} 
\usepackage{dsfont}

\usepackage[normalem]{ulem}

\begin{document}

\title{Capacity of multimode quantum Gaussian channels}

\author{Maria Popławska, Marcin Jarzyna
        % <-this % stops a space
\thanks{Maria Popławska is with the Centre for Quantum Optical Technologies, Centre
of New Technologies, University of Warsaw, Banacha 2c, 02-097 Warszawa, Poland (e-mail: m.poplawska@cent.uw.edu.pl)

Marcin Jarzyna is with the Centre for Quantum Optical Technologies, Centre
of New Technologies, University of Warsaw, Banacha 2c, 02-097 Warszawa, Poland (e-mail: m.jarzyna@cent.uw.edu.pl)}% <-this % stops a space
%\thanks{Manuscript received xxx; revised xxx}
}

%<-- added by MP% \markboth{Journal of \LaTeX\ Class Files,~Vol.~14, No.~8, August~2021}%
%<-- added by MP%{Shell \MakeLowercase{\textit{et al.}}: A Sample Article Using IEEEtran.cls for IEEE Journals}
%<-- added by MP%\IEEEpubid{0000--0000/00\$00.00~\copyright~2021 IEEE}

\maketitle

\begin{abstract} 
We derive explicit formulas for the capacity of multimode quantum Gaussian channels which serve as a fundamental model for optical version of multiple-input multiple-output channels. We show that increasing the number of modes under fixed power constraint does not reduce the capacity. We derive an analytical formula for the ensemble-averaged Holevo capacity in the case of random passive transformations. The analogous results are also obtained for capacities achievable under homodyne and heterodyne detection. We further discuss the generalization of the model to include weak active transformations.
\end{abstract}

\begin{IEEEkeywords}
Gaussian channels, MIMO, channel capacity, quantum channel, quantum communication, optical modulation
\end{IEEEkeywords}

\section{Introduction} 
\IEEEPARstart{O}{ptical} communication is a key technology for high-capacity information transmission \cite{Essiambre2012}. Platforms ranging from free-space optical links to modern optical fiber networks enable reliable communication over long distances and at unprecedented data rates \cite{Hout2025}.
A key challenge in modern communication is to increase channel capacity without proportionally increasing physical resources such as bandwidth or power.

A major breakthrough in radio frequency communications was the development of multiple-input multiple-output (MIMO) systems, which employ multiple transmitting and receiving antennas to significantly increase link capacity \cite{Telatar1999}. 
Analogues of MIMO have since been introduced in optical communications, most notably in the form of Spatial-Division Multiplexing \cite{Richardson2013} and Mode-Division Multiplexing \cite{Ryf2012}, which exploit multiple spatial or modal degrees of freedom of the optical field. These approaches are typically formulated within a classical framework, where  the signal is treated as a continuous waveform \cite{Proakis2008}. However, at the fundamental level light is described by the laws of quantum mechanics which on one hand opens new possibilities, such as quantum enhanced measurement strategies, but on the other imposes restrictions e.g. in the form of unavoidable quantum noise. This naturally raises the problem of quantum description of optical MIMO channels, similarly as has been already achieved for single mode case \cite{Giovannetti2004,Giovannetti2014, Shapiro2009}.

Within quantum information theory, optical communication links are commonly modelled as bosonic Gaussian quantum channels \cite{Weedbrook2012, Eisert2007}. This framework reflects the widespread experimental use of Gaussian states, such as coherent states generated by laser sources or thermal states, and accommodates the description of typical noise effects as well as multimode transmission. Crucially, using tools from quantum information theory, one can characterize the fundamental limits on information transmission rates in terms of the Holevo bound \cite{Holevo1973}. Unlike capacities derived under specific classical detection strategies, such as conventional Shannon-Hartley bound, the Holevo limit provides an upper limit on the accessible classical information for a given quantum channel and input ensemble, corresponding to the optimal quantum measurement strategy \cite{Banaszek2020}. This allows one to quantify the ultimate performance of optical systems independently of the receiver architecture.

The purpose of this work is to extend the fundamental ideas introduced in~\cite{Telatar1999}, concerning the role of the number of modes and random scattering in the information capacity scaling, to a quantum channel description of optical MIMO communication systems. Specifically, we show that for a channel governed by passive or phase-insensitive operations, one can choose an appropriate encoding basis that decomposes the multimode channel into parallel subchannels. The Holevo information can then be maximized by applying the quantum analogue of the water-filling algorithm \cite{Schafer2009} to a coherent state modulation in the encoding basis. Under a given input power constraint, the channel capacity is non-decreasing with number of modes used for communication. Moreover, we show that for random passive transformations, which are analogous to random signal scattering in radio communication, the expected information capacity is given by an analytical formula derived with the ensemble-averaged spectral density which differs from the classical case.

This paper is organized as follows. The mathematical description of multimode optical communication channels in terms of bosonic Gaussian quantum channels is presented in Sec.~\ref{sec:channels}. In Sec.~\ref{sec:c_quant} we derive the capacity for general Gaussian multimode channels and give simplified expressions for passive and phase-insensitive case. Next, random scattering of the signal is modelled as a random passive transformation within the quantum channel formalism and an analytical expression for the expected information capacity is derived in Sec.~\ref{sec:random}. An extension to weak active transformations is also discussed there. Finally, Sec.~\ref{sec:Conclusions} concludes the paper.

\section{Classical and Quantum Multimode Channel}
\label{sec:channels}
\begin{figure}[!t]
\centering
\includegraphics[width=\columnwidth]{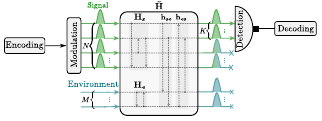}
\caption{Scheme of a communication link involving a multimode Gaussian channel. The transmitter prepares $N$ signal modes with Gaussian modulation and the receiver receives $K$ modes while the environment consists of $M$ modes that are beyond control by either the transmitter or receiver. The quantum channel couples signal and environmental degrees of freedom through matrices $\mathbf{h}_{\textrm{se}}$ and $\mathbf{h}_{\textrm{es}}$, resulting in leakage of the signal energy into the environment. The receiver performs possible collective measurements on the output signal modes.}
\label{fig:communication_scheme}
\end{figure}

\subsection{Classical communication}
\label{sec:channels_classical}
A basic model for a classical multimode optical channel, presented schematically in Fig.~\ref{fig:communication_scheme} consists of a transformation $\mathbf{H}_{\textrm{s}}$ acting on $N$ optical input modes and returning $K$ output ones. The transmitter and receiver can perform in general correlated modulation and detection across all modes respectively. An optical field in the i-th mode is characterized by two orthogonal quadratures $q^{\textrm{s}}_i$ and $p^{\textrm{s}}_i$ composing a $2N$-dimensional signal quadrature vector $\mathbf{r}^{\textrm{s}}=(q^{\textrm{s}}_1,\dots, q^{\textrm{s}}_N,p^{\textrm{s}}_1,\dots, p^{\textrm{s}}_N)^{\text{T}}$ which transforms under the multimode channel as $\mathbf{r}^{\textrm{s}}_{\textrm{out}}=\mathbf{H}_{\textrm{s}}\mathbf{r}^{\textrm{s}}_{\textrm{in}}+\mathbf{n}$, where $\mathbf{r}^{\textrm{s}}_{\textrm{in}}$ and $\mathbf{r}^{\textrm{s}}_{\textrm{out}}$ denote input and output quadrature vectors respectively and $\mathbf{n}$ is an additional displacement that may introduce random noise. The signal is characterized by the covariance matrix $\mathbf{V}^{\textrm{s}}=\langle \mathbf{r}^{\textrm{s}}(\mathbf{r}^{\textrm{s}})^{\textrm{T}}\rangle-\langle\mathbf{r}^{\textrm{s}}\rangle\langle (\mathbf{r}^{\textrm{s}})^{\textrm{T}}\rangle$ which in turn transforms as $\mathbf{V}^{\textrm{s}}_{\textrm{out}}=\mathbf{H}_{\textrm{s}}\mathbf{V}^{\textrm{s}}_{\textrm{in}}\mathbf{H}_{\textrm{s}}^{\textrm{T}}+\xi\mathds{1}_{2K}$, where $\mathbf{V}^{\textrm{s}}_{\textrm{in}}$ and $\mathbf{V}^{\textrm{s}}_{\textrm{out}}$ denote input and output signal covariance matrices and $\xi=\langle \mathbf{n}\mathbf{n}^\textrm{T}\rangle$ represents uniform additive noise that may be additionally introduced by the channel. With the above assumptions, it was shown in \cite{Telatar1999} that one may calculate multimode channel capacity by noticing that under singular value decomposition (SVD) of the transformation matrix $\mathbf{H}_{\textrm{s}}=\mathbf{U}\mathbf{D}\mathbf{W}^{\textrm{T}}$, where $\mathbf{U},\mathbf{W}$ are respectively $K\times K$ and $N\times N$ unitary matrices and $\mathbf{D}$ is diagonal rectangular $K\times N$ matrix containing singular values, the covariance matrix may be written as
\begin{equation}
    \tilde{\mathbf{V}}^{\textrm{s}}_{\textrm{out}}=\mathbf{D}\tilde{\mathbf{V}}^{\textrm{s}}_{\textrm{in}}\mathbf{D}+\xi\mathds{1}_{2K},
\end{equation}
where $\tilde{\mathbf{V}}^{\textrm{s}}_{\textrm{out}}=\mathbf{U}^{\textrm{T}}\mathbf{V}^{\textrm{s}}_{\textrm{out}}\mathbf{U}$ and $\tilde{\mathbf{V}}^{\textrm{s}}_{\textrm{in}}=\mathbf{W}^{\textrm{T}}\mathbf{V}^{\textrm{s}}_{\textrm{in}}\mathbf{W}$ denote output and input covariance matrices rotated to appropriate mode bases given by $\mathbf{U}$ and $\mathbf{W}$. Importantly, the rotation transformations $\mathbf{U}$ and $\mathbf{W}$ can be applied by the receiver and transmitter either by physical signal manipulation or by postprocessing. The multimode channel capacity under total power constraint $P$ can be then written as
\begin{equation}\label{eq:telatar}
    C(P)=\max_{\{P_k\}}\sum_{k=1}^K\log_2\left(1+\frac{\lambda_k P_k}{\xi}\right),
\end{equation}
where $\lambda_k=\mathbf{D}_{kk}^2$ denotes square of the singular value of $\mathbf{H}_{\textrm{s}}$, $P_k$ is the power sent to the $k$-th singular mode and the maximization is performed over different power distributions that satisfy $\sum_{k=1}^K P_k\leq P$. The latter can be done by using water-filling method \cite{Telatar1999}, resulting in the formula
\begin{equation}
C(P)=\sum_{k=1}^K\left\{\log_2 \frac{\mu\lambda_j}{\xi}\right\}^{+},\quad P=\sum_{k=1}^K\left\{\mu-\frac{\xi}{\lambda_j}\right\}^{+},    
\end{equation}
where $\mu$ is a "water-level" coefficient to be determined from the power constraint on the right and $\{x\}^{+}=\max(x,0)$. Crucially, note that the capacity in Eq.~(\ref{eq:telatar}) is a sum of standard expressions resulting from applying the well known Shannon-Hartley bound to each singular mode. In particular it depends only on the signal to noise ratios in each mode.

\subsection{Multimode Gaussian quantum channels and the Holevo limit}
\label{sec:channels_quantum}
In quantum mechanical picture the quadratures become operators and the optical field is described by quantum states \cite{Banaszek2020}. A standard optical channel is described by a quantum bosonic Gaussian channel \cite{Weedbrook2012, Braunstein_2005}, which is characterized by the fact that it transforms Gaussian states into Gaussian states. The latter is a class of states which can be completely characterized by first and second quadrature moments and include such examples like coherent states and squeezed states \cite{Weedbrook2012}. Gaussian channels contain such important transformations as losses, additive noise, phase sensitive and insensitive amplification and many others. Importantly, the capacity of single mode Gaussian quantum channels has been found \cite{Giovannetti2004,Giovannetti2014,Schafer2016} and it is known that in the phase insensitive scenario it can be saturated by a Gaussian ensemble of coherent states.

It is instructive to consider the action of general Gaussian channel including not only the signal modes but also the environment, as seen in Fig.~\ref{fig:communication_scheme}. We consider an input signal and environment consisting of $N$ and $M$ modes respectively. The vector of first moments of quadratures and the combined covariance matrix may be then decomposed as
\begin{subequations}
\begin{align}
\mathbf{r}_{\mathrm{in}} &=
(q^{\textrm{s}}_{1},\dots,q^{\textrm{s}}_{N},p^{\textrm{s}}_{1},\dots,p^{\textrm{s}}_{N},
q^{\textrm{e}}_{1},\dots,q^{\textrm{e}}_{M},p^{\textrm{e}}_{1},\dots,p^{\textrm{e}}_{M})^{\mathrm{T}}, \\
\mathbf{V}_{\mathrm{in}} &= \mathbf{V}^{\textrm{s}}_{\mathrm{in}} \oplus \mathbf{V}^{\textrm{e}}_{\mathrm{in}},
\end{align}
\end{subequations}
where the superscripts $\textrm{s}$ and $\textrm{e}$ denote signal and environment subsystems respectively. The action of the Gaussian channel is described by a symplectic transformation $\tilde{\mathbf{H}}$ which can be decomposed as
\begin{equation}
\label{eq:tilde_H_definition}
\tilde{\mathbf{H}} =
\begin{bmatrix}
\mathbf{H}_{\textrm{s}} & \mathbf{h}_{\textrm{se}} \\
\mathbf{h}_{\textrm{es}} & \mathbf{H}_{\textrm{e}}
\end{bmatrix}.
\end{equation}
Here, the subblocks $\mathbf{h}_{\textrm{se}}$ and $\mathbf{h}_{\textrm{es}}$ describe the coupling between signal and environmental degrees of freedom and $\mathbf{H}_e$ characterizes transformation of the environment. Importantly, $\tilde{\mathbf{H}}$ must satisfy the relation $\tilde{\mathbf{H}}\Omega\tilde{\mathbf{H}}^{\mathrm{T}}=\Omega$, where $\Omega$ is the symplectic form on a $2(N+M)$ dimensional phase-space:
\begin{equation}
\Omega= \Omega_{2N} \oplus \Omega_{2M}, \quad \Omega_{2K}=
\begin{bmatrix}
0 & \mathds{1}_{K} \\
-\mathds{1}_{K} & 0
\end{bmatrix}.
\end{equation}
At the receiver, only $K$ of the signal modes are accessible. The resulting output state of the signal subsystem is therefore described by
\begin{subequations}
\label{eq:output_state}
\begin{align}
\mathbf{r}_{\mathrm{out}} &= \mathbf{H}_{\textrm{s}}\mathbf{r}^{\textrm{s}}_{\mathrm{in}}
+ \mathbf{h}_{\textrm{se}}\mathbf{r}^{\textrm{e}}_{\mathrm{in}}
= \mathbf{H}_{\textrm{s}}\mathbf{r}^{\textrm{s}}_{\mathrm{in}} + \mathbf{d}, \\
\mathbf{V}^{\textrm{s}}_{\mathrm{out}} &=
\mathbf{H}_{\textrm{s}}\mathbf{V}^{\textrm{s}}_{\mathrm{in}}\mathbf{H}_{\textrm{s}}^{\mathrm{T}}
+ \mathbf{h}_{\textrm{se}}\mathbf{V}^{\textrm{e}}_{\mathrm{in}}\mathbf{h}_{\textrm{se}}^{\mathrm{T}}
= \mathbf{H}_{\textrm{s}}\mathbf{V}^{s}_{\mathrm{in}}\mathbf{H}_{\textrm{s}}^{\mathrm{T}} + \mathbf{Y},
\end{align}
\end{subequations}
where $\mathbf{H}_{\textrm{s}}$ describes, similarly as before, the effective linear transformation of the signal and $\mathbf{Y}=\mathbf{h}_{\textrm{se}}\mathbf{V}_{\textrm{in}}^{\textrm{e}}\mathbf{h}_{\textrm{se}}^{\mathrm{T}}$ represents the noise induced by coupling to the environment while $\mathbf{d}=\mathbf{h}_{\textrm{se}}\mathbf{r}_{\textrm{in}}^{\textrm{e}}$ is the noise induced displacement.
An important class of Gaussian quantum channels are minimal noise channels. Unlike in the classical case, where the additive noise $\xi$ could take any value, including $0$, the matrices describing a valid quantum channel must satisfy
\begin{equation}
\label{eq:minimal_noise_channel}
    \mathbf{Y}\geq \frac{i}{2}\mathbf{\Sigma},\quad\mathbf{\Sigma}=\Omega_{2K}-\mathbf{H}_{\textrm{s}}\Omega_{2N}\mathbf{H}_{\textrm{s}}^{\textrm{T}},
\end{equation}
where $\Omega_{2K}$ is the symplectic form on the $2K$ dimensional output subspace.
One may show that for a given $\mathbf{H}_s$ the minimal noise condition is satisfied by a noise matrix
\begin{equation}\label{eq:min_noise}
    \mathbf{Y}=\frac{1}{2}\left|\mathbf{\Sigma}\right|=\frac{1}{2}\left|\Omega_{2K}-\mathbf{H}_s\Omega_{2N}\mathbf{H}_s^{\textrm{T}}\right|.
\end{equation}

To further simplify Eq.~\eqref{eq:min_noise} we will assume that the signal transformation matrix adopts a block form
\begin{equation}
\label{eq:H_s_phase_insensitive_passive}
    \mathbf{H}_{\textrm{s}} = \begin{bmatrix}
        \mathbf{H}_{1} & -\mathbf{H}_{2}\\
        \mathbf{H}_{2} & \mathbf{H}_{1}
    \end{bmatrix},
\end{equation}
where $\mathbf{H}_{1}, \mathbf{H}_{2}$ are $K\times N$ real matrices. This is the case for a wide range of practical scenarios, in particular, matrices of the form in Eq.~\eqref{eq:H_s_phase_insensitive_passive} describe general passive transformations as well as general phase-insensitive transformations e.g. beam-splitting, phase shifts or phase-insensitive amplification. Note, that more general, phase-sensitive Gaussian operations such as single-mode squeezing or phase sensitive amplification are not covered by Eq.~\eqref{eq:H_s_phase_insensitive_passive}. Because matrices defined in Eq.~\eqref{eq:H_s_phase_insensitive_passive} satisfy $\Omega_{2K}\mathbf{H}_{\textrm{s}} = \mathbf{H}_\textrm{s}\Omega_{2N}$ \cite{Idel2016OnQA}, one can rewrite Eq.~\eqref{eq:min_noise} using this property and obtain 
\begin{equation}\label{eq:Y}
    \mathbf{Y}=\frac{1}{2}\left|\mathds{1}_{2K}-\mathbf{H}_{\textrm{s}}\mathbf{H}_{\textrm{s}}^{\textrm{T}}\right|.
\end{equation}
The noise in \eqref{eq:Y} is attained for environment modes in the vacuum state. One can easily model noise introduced by environmental modes in a uniform thermal state with $n$ noisy photons per mode and additive noise $\xi$ as
\begin{equation}\label{eq:thermal}
    \mathbf{Y}=\left(n+\frac{1}{2}\right)\left|\mathds{1}_{2K}-\mathbf{H}_{\textrm{s}}\mathbf{H}_{\textrm{s}}^{\textrm{T}}\right|+\xi\mathds{1}_{2K}.
\end{equation}
Note, however, that for channels that are not described by Eq.~\eqref{eq:H_s_phase_insensitive_passive} one needs to use the general form of minimal-noise matrix in Eq.~\eqref{eq:min_noise}. 

\section{Capacity of a quantum multimode channel}
\label{sec:c_quant}

In contrast to classical scenario, where the capacity can be found from the Shannon-Hartley formula, in the quantum mechanical description one needs to use the Holevo bound \cite{Holevo1973}. The latter characterizes the maximal classical information rate that can be transmitted through a channel using arbitrary quantum measurements, including collective ones on many channel repetitions. For a general ensemble of states $\{\rho_x\}$ appearing with probability $p_x$ and sent through a general quantum channel $\Phi$, the Holevo bound is given by
\begin{equation}\label{eq:holevo}
\chi = S\!\left[\sum_x p_x \Phi(\rho_x)\right]
      - \sum_x p_x S\!\left[\Phi(\rho_x)\right],
\end{equation}
where $S(\rho)=-\mathrm{Tr}(\rho\log\rho)$ is the von Neumann entropy of a state $\rho$. The Holevo information optimized over all ensembles of input states under a given power constraint is the maximum theoretically reachable information transmission rate of a quantum channel and is known as classical capacity of a quantum channel. Eq.~(\ref{eq:holevo}) usually results in cumbersome expressions, however, fortunately, for Gaussian ensembles transmitted across Gaussian channels the Holevo bound admits a closed-form formula. This is because the von Neumann entropy of a $K$-mode Gaussian state is specified only by the symplectic eigenvalues $\nu_k$ of its covariance matrix, i.e. $S(\rho)=\sum_{k=1}^Kg(\nu_k-\frac{1}{2})$, where $g(x) = (x+1)\log_2(x+1)-x\log_2(x)$ and the symplectic eigenvalues of a covariance matrix $\mathbf{V}$ are given by positive eigenvalues of $i\Omega_{2K}\mathbf{V}$.

We assume below a Gaussian amplitude modulation of coherent states. In such case covariance matrix for the output state is independent of the modulation amplitude $\alpha$, i.e. $V_{\textrm{out}}^{\textrm{s}}|_\alpha=\mathbf{H}_s\mathbf{H}_s^{\textrm{T}}/2+\mathbf{Y}$, therefore, also respective symplectic eigenvalues $\{\nu_k\}$ do not change \cite{Pilyavets2012} and the Holevo information is given by 
\begin{equation}
\label{eq:holevo_by_symplectic_values}
    \chi=\sum_{k=1}^K\left[g\left(\bar{\nu}_k-\frac{1}{2}\right)-g\left(\nu_k-\frac{1}{2}\right)\right],
\end{equation}
where $\bar{\nu}_k$ are symplectic eigenvalues of the covariance matrix $\bar{\mathbf{V}}_{\textrm{out}}^{\textrm{s}}=\mathbf{H}_{\textrm{s}}\bar{\mathbf{V}}_{\textrm{in}}^{\textrm{s}}\mathbf{H}_{\textrm{s}}^{\textrm{T}}+\mathbf{Y}$ of the output state when the input is the averaged state with covariance matrix $\bar{\mathbf{V}}_{\textrm{in}}^{\textrm{s}}$.

\begin{figure*}
\centering
\includegraphics[width=\linewidth]{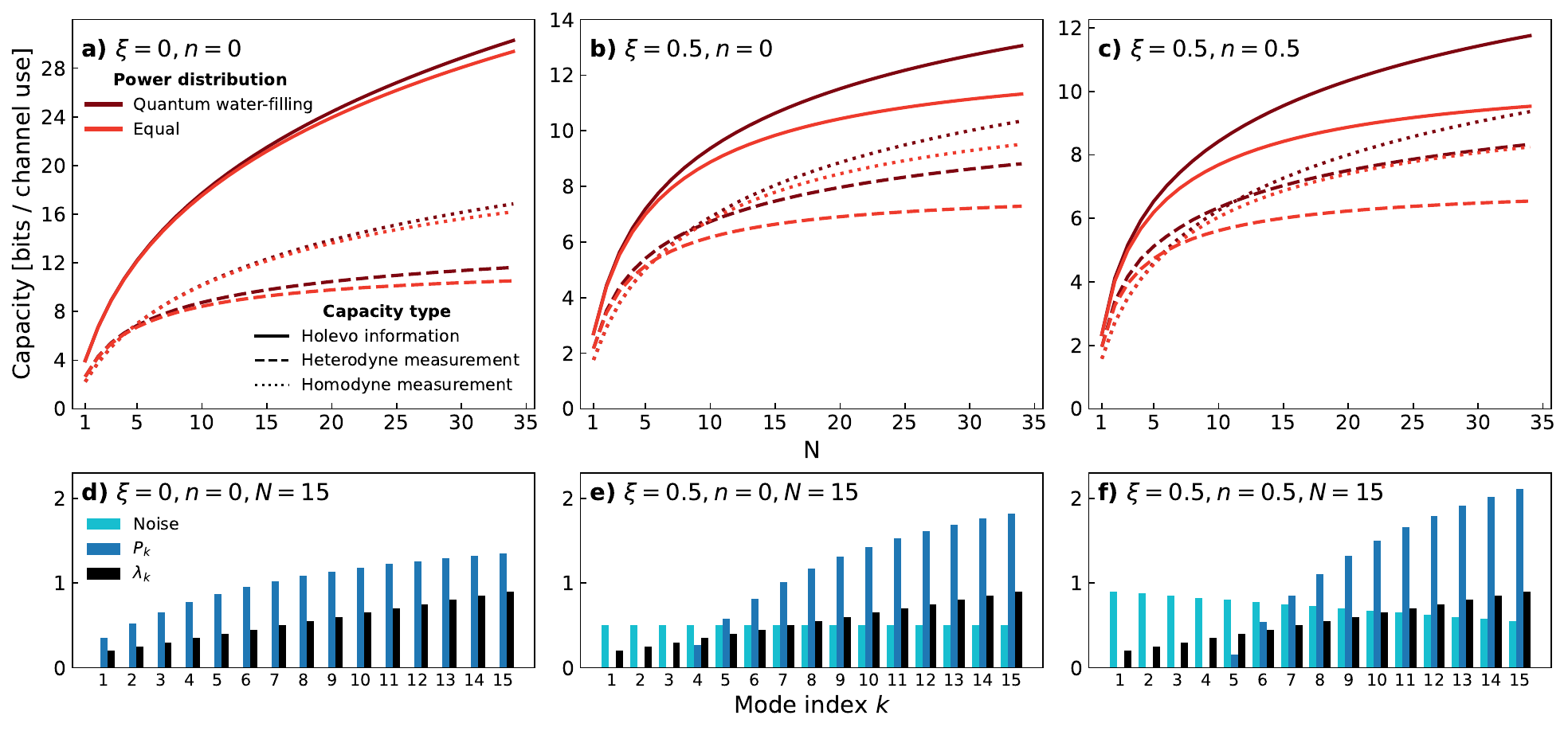}
\caption{Holevo channel capacity as well as capacity of homodyne and heterodyne receiver for uniform and optimal power distribution as a function of number of modes assuming $N=K$ for input power given by average number of photons $P=15$ for the noiseless scenario (a), additive noise (b) and both additive and channel noise (c). The transmission coefficients of singular modes are taken as $\lambda_k=0.2+0.7\times k/N\in [0.2,0.9]$, $k=1,\dots, N$. The corresponding optimal distribution of power between different modes for the Holevo capacity for $N=15$ with singular modes transmission and noise indicated (d-f).}
\label{fig:capacity}
\end{figure*}

One can further simplify the formula for Holevo bound by assuming uniform thermal noise in the environment, Eq.~\eqref{eq:thermal} and decomposing the transformation matrix. Assuming $\mathbf{H}_s$ is of the form in Eq.~\eqref{eq:H_s_phase_insensitive_passive}, i.e. phase-insensitive or passive, using the SVD decomposition $\mathbf{H}_{s}=\mathbf{UDW}^{\mathrm{T}}$, the output covariance matrix can be written in the rotated basis as
\begin{equation}\label{eq:diagonal}
\tilde{\mathbf{V}}^{s}_{\mathrm{out}} = 
\mathbf{D}\tilde{\mathbf{V}}^{s}_{\mathrm{in}}\mathbf{D}
+ \left(n+\frac{1}{2}\right)\left|\mathds{1}_{2K}-\mathbf{D}^2\right| + \xi\mathds{1}_{2K},
\end{equation}
where $\tilde{\mathbf{V}}^{\textrm{s}}_{\textrm{out}} = \mathbf{U}^{\text{T}} V^{\textrm{s}}_{\textrm{out}}\mathbf{U}$, $\tilde{\mathbf{V}}^{\textrm{s}}_{\textrm{in}} = \mathbf{W}^{\text{T}}V^{\textrm{s}}_{\textrm{in}}\mathbf{W}$. Crucially, for matrices given by Eq.~\eqref{eq:H_s_phase_insensitive_passive} the orthogonal matrices $\mathbf{U,W}$ in SVD are also symplectic (see Appendix \ref{sec:appendix_SVD_for_optical_transformations}) and therefore represent valid passive optical transformations. If one chooses the input state in such a basis that $\mathbf{\tilde{V}}^{\textrm{s}}_{\textrm{in}}$ is diagonal, then also the $ \tilde{\mathbf{V}}^{\textrm{s}}_{\textrm{out}}$ must be diagonal. Using then Eq.~\eqref{eq:diagonal}, the  Holevo information can be written explicitly in terms of the input covariance matrix elements for the averaged state $(\tilde{V}_{\textrm{in}}^{\textrm{s}})_{kk}=P_k+1/2$ and square of singular values $\lambda_k$ of 
$\mathbf{H}_{\textrm{s}}$, i.e. $\lambda_k=\mathbf{D}_{kk}^2$
as 
\begin{multline}
\label{eq:holevo_multi}
\chi = \sum^{K}_{k=1}\left\{ g\left[\lambda_k P_k+ \tfrac{\lambda_k-1}{2}+\left(n+\tfrac{1}{2}\right)\left|1-\lambda_{k}\right| +  \xi\right] \right.+\\
\left.-g\left[\tfrac{\lambda_k-1}{2}+\left(n+\tfrac{1}{2}\right)\left|1-\lambda_{k}\right| + \xi\right] \right\},
\end{multline}
where $P_k$ denotes the power devoted to $k$-th input singular mode and we used the fact that input covariance matrix for coherent states is equal to $\mathds{1}$/2.

Crucially, note that since the matrix $\mathbf{D}$ is diagonal, the effective multimode Gaussian channel in Eq.~\eqref{eq:diagonal} is in fact a collection of independent, i.e. uncorrelated one-mode Gaussian channels. Because the classical capacity of such quantum channels is additive and for each of them achieved by Gaussian ensemble of coherent states \cite{Hiroshima2006, Giovannetti2014}, the Holevo information in Eq.~\eqref{eq:holevo_multi}, which is just a sum of Holevo informations for individual singular modes under coherent state modulation, is also the classical capacity of the multimode quantum Gaussian channel under a particular power distribution $\{P_k\}$. 
The global multimode capacity under total power constraint $P$ is then equal to Eq.~\eqref{eq:holevo_multi} optimized over power distribution between the singular modes. 
The optimal power allocation can be found by using water-filling strategy \cite{Telatar1999, Schafer2012}, analogously to the classical case, with an important caveat that the minimal noise experienced by each mode depends on the signal transformation matrix, resulting in
\begin{multline}\label{eq:capacity_optimal}
    C(P)=\sum_{k=1}^K\left\{g\left(\frac{1}{\mu^{1/\lambda_k}-1}\right)+\right.\\\left.-g\left[\frac{\lambda_k-1}{2}+\left(n+\frac{1}{2}\right)\left|1-\lambda_k\right|+\xi\right]\right\}^{+},
\end{multline}
where $\mu$ is a coefficient that has to be determined from the power constraint
\begin{equation}
  P=\sum_{k=1}^K\frac{1}{\lambda_k}\left\{\frac{1}{\mu^{1/\lambda_k}-1}-\frac{\lambda_k-1}{2}-\left(n+\frac{1}{2}\right)\left|1-\lambda_k\right|-\xi\right\}^{+},
\end{equation} and symmetric modulation of quadratures in every mode is assumed.

Importantly, note that one can perform physically valid SVD only for transformations taking the form in Eq.~\eqref{eq:H_s_phase_insensitive_passive}. However, unlike in the classical picture, for general transformations the resulting SVD unitary matrices $\mathbf{U}$ and $\mathbf{W}$ are not symplectic, which is a necessary condition for valid Gaussian quantum transformations. In such cases one can still show that there exist a canonical form of the transformation matrix \cite{Wolf_2008,Caruso2008}, however, it is not strictly diagonal but rather involves not necessarily diagonal Jordan blocks. The latter is the case for general active transformations and in such instances one needs to use the full formula Eq.~\eqref{eq:min_noise}. Similarly, note that even if the resulting SVD matrices are physical, one in general cannot reduce the covariance matrix to the form in Eq.~\eqref{eq:diagonal} and, consequently, use Eqs.~\eqref{eq:holevo_multi} and~\eqref{eq:capacity_optimal}, if the additive noise or thermal state of the environment are not uniform across the modes.  In order to calculate Holevo information in such cases one needs to use the initial, most general formula for Gaussian states Eq.~\eqref{eq:holevo_by_symplectic_values}.

The capacity formula from Eq.~\eqref{eq:holevo_multi} gives capacity optimized over all possible ensembles of quantum states and all reception strategies allowed by laws of quantum mechanics, some of which may be potentially very difficult to implement in practice.  A standard detection strategy in optical communication is heterodyne or homodyne measurement. The capacities achievable with such receivers are given by the classical mutual information of the corresponding Gaussian measurement outcomes, which can be expressed in a form analogous to Eq.~\eqref{eq:telatar}, with appropriately defined signal and noise covariance matrices, see Appendix~\ref{sec:appendix_hetero_homodyne_measurements}.

It is seen in Fig.~\ref{fig:capacity}(a-c) that increasing the number of input/output modes leads to a non-decreasing capacity. However, in the noisy case, Fig.~\ref{fig:capacity}(b),(c), the capacity saturates at some value, whereas in the noiseless scenario, Fig.~\ref{fig:capacity}(a), it slowly grows to infinity. Note also, that for small number of modes it is beneficial to use heterodyne detection, whereas for large number of modes homodyne receiver performs better, although both schemes do not attain the quantum bound. It is seen in Fig.~\ref{fig:capacity}(d-f) that the optimal power distribution crucially depends on the type of noise present in the channel. In the noiseless scenario, Fig.~\ref{fig:capacity}(d), the power is unequally distributed in all singular modes, depending on the transmission. In contrast, in the noisy case, it is optimal to not modulate light in modes for which the ratio of noise and transmission is too large.  This is especially crucial if the environmental noise is present, since in such case the effective noise experienced by each singular mode depends on its transmission.

\section{Random Optical Quantum Gaussian Channels}
\label{sec:random}
In this section, we introduce a systematic framework for modelling the propagation of optical signals through randomly scattering media within the formalism of quantum Gaussian channels. This is an equivalent of a classical Rayleigh fading channel describing fluctuations of the received signal amplitude caused by e.g. multipath propagation in the environment with many scatterers and whose capacity was derived in \cite{Telatar1999}. Importantly, note that this model does not include effects like correlated-scattering and, similarly to the classical Rayleigh-fading, treats all modes as statistically equivalent.
We also assume no channel state information at the transmitter, i.e., the transmitter knows only the statistical distribution of $\mathbf{H}_{\mathrm{s}}$ but not its instantaneous realization.

Using formalism of Bogoliubov transformations \cite{Braunstein_2005} any symplectic transformation in phase space can be represented as
\begin{equation}
\label{eq:tilde_H_definition_A_B1}
    \Tilde{\mathbf{H}} = \begin{bmatrix}
        \textrm{Re}{\mathbf{A}} + \textrm{Re}{\mathbf{B}} &  -\textrm{Im}{\mathbf{A}} + \textrm{Im}{\mathbf{B}} \\
        \textrm{Im}{\mathbf{A}} + \textrm{Im}{\mathbf{B}} & \textrm{Re}{\mathbf{A}} - \textrm{Re}{\mathbf{B}}  
    \end{bmatrix},
\end{equation}
for quadrature ordering: $\mathbf{r}_{in} =(q_{1},\dots,q_{N+M},p_{1},\dots,p_{N+M})^{\text{T}}$, where $\mathbf{A}, \mathbf{B}$ are real matrices satisfying $\mathbf{AB}^{\textrm{T}} = \mathbf{BA}^{\textrm{T}}$, $\mathbf{AA}^{\dagger} = \mathbf{BB}^{\dagger}+\mathds{1}$ \cite{Weedbrook2012}.
One can use convenient parametrization to separate passive and active parts of transformation
\begin{subequations}
\label{eq:A_B_definitions}
\begin{align}
    \mathbf{A} &= \mathbf{U}_1 \cosh{\mathbf{R}}\mathbf{U}_2,\\
    \mathbf{B} &=\mathbf{U}_1 \sinh{\mathbf{R}}\mathbf{U}_2^{*},
\end{align}
\end{subequations}
where $\mathbf{U_1}, \mathbf{U_2}$ are $(N+M)\times (N+M)$ unitary matrices and $\mathbf{R}$ is a real diagonal matrix that describe a possible squeezing of optical quadratures introduced by the transformation.

\subsection{Passive optical transformations}
\label{sec:passive}

Scattering processes that redistribute optical power among different optical modes and induce random phase shifts can be represented as random passive linear transformations acting on the field operators. In terms of transformation in Eq.~\eqref{eq:A_B_definitions} this corresponds to $\mathbf{R}=0$, that is
\begin{equation}
    \mathbf{A} = \mathbf{U}, \quad{} \mathbf{B} =0,
\end{equation}
where $\mathbf{U}$ is a unitary transformation. Within this model, the resulting channel is fully characterized by a random transformation matrix, allowing the ensemble-averaged information capacity to be analyzed analytically. Similarly as above, we consider a symmetric Gaussian ensemble of coherent states for information encoding, which is optimal for passive channels. Because the Holevo information in Eq.~\eqref{eq:holevo_multi} as well as homodyne and heterodyne measurement capacities for diagonalized multimode channel have a form of a sum of terms depending on individual singular values, one can calculate the expected value of the channel capacity through integration as 
\begin{equation}
\label{eq:expected_value_capacity_RMT}
	C=\mathbb{E} \left[\sum^{K}_{j=1}C(\lambda_j)\right] = \sum^{K}_{j=1}\mathbb{E} \left[C(\lambda_j)\right] = K \int C(\lambda) p_{\lambda}(\lambda)\textrm{d}\lambda
\end{equation}
where $p_{\lambda}(\lambda)$ is the ensemble-averaged spectral density \cite{Livan2011} and $C(\lambda)$ denotes the capacity contribution associated with an eigenvalue $\lambda$ of the matrix $\mathbf{H}_{\textrm{s}}\mathbf{H}_{\textrm{s}}^{\text{T}}$, where matrix $\mathbf{H}_{\textrm{s}}$ describes the transformation on the signal modes. The quantity $C(\lambda)$ is the appropriate, i.e. homodyne, heterodyne or Holevo, information capacity of a single-mode lossy bosonic channel with transmission coefficient $\lambda$, assuming the same additive noise $\xi$ and average number of thermal photons $n$ in the environment. Importantly, since the transmitter does not posses any knowledge about a particular realization of the channel transformation $\mathbf{H}_s$ and the channel statistical ensemble is independent under mode permutations, it is optimal to use a uniform input power distribution without the need of waterfilling. Consider now a transformation matrix $\mathbf{A}$ with dimensions $(N+M) \times (N+M)$ to be a random unitary sampled from Haar measure, that acts both on the signal and environmental degrees of freedom. It can be divided into subblocks to differentiate those two types of modes
\begin{equation}\label{eq:amatrix}
    \mathbf{A} = \begin{bmatrix}
        \mathbf{A}_{1} & \mathbf{A}_{2} \\
        \mathbf{A}_{3} & \mathbf{A}_{4}
    \end{bmatrix},
\end{equation}
where $\dim{\mathbf{A}_{1}}= K\times N$. 
Using definitions from Eq.~\eqref{eq:tilde_H_definition} and Eq.~\eqref{eq:tilde_H_definition_A_B1} one can write: 
\begin{equation}
\label{eq:H_s_passive_transformation}
    \mathbf{H}_{\textrm{s}} = 
    \begin{bmatrix}
        \textrm{Re}{\mathbf{A}_{1}} & -\textrm{Im}{\mathbf{A}_{1}}\\
        \textrm{Im}{\mathbf{A}_{1}} &  \textrm{Re}{\mathbf{A}_{1}}\\
    \end{bmatrix}.
\end{equation}

\begin{figure}
\centering
\includegraphics[width=\columnwidth]{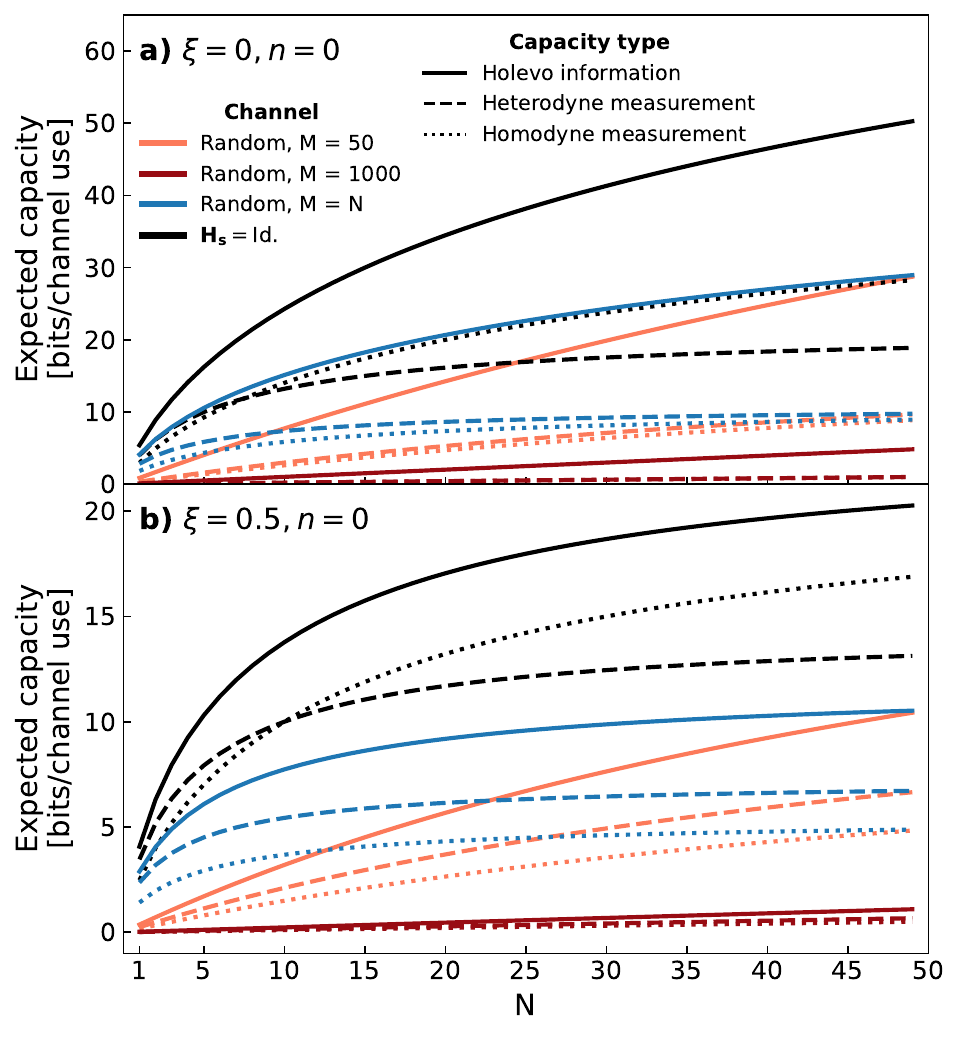}
\caption{Capacity of a random passive optical channel as a function of the number of signal transmitter modes $N$. The total input power $P=15$ is distributed equally among the $N$ modes, and the number of receiver modes is $K=N$.
The black curves represent the information capacity of the identity channel with the same parameters $\xi,n$.} 
\label{fig_expected_capacities_RMT}
\end{figure}

\begin{figure*}
\centering
\includegraphics[width=\linewidth]{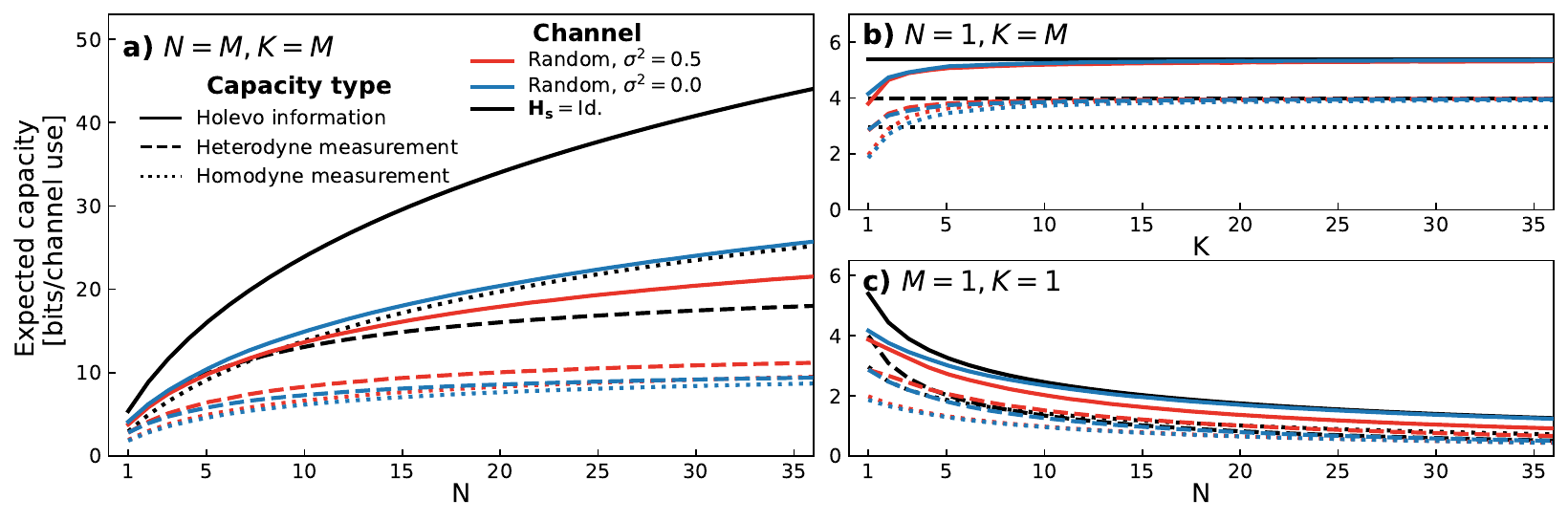}
\caption{Capacity of a random weakly active optical channel for different values of $\sigma^{2}$, the variance used to sample squeezing parameters, for total input power $P=15$, for $N=K=M$ (a), $N=1,K=M$ (b) and $M=K=1$ (c) as a function of the number of  modes for uniform power distribution among signal modes and $\xi=0, n=0$.
For each number of modes, capacities were obtained from Monte Carlo averages over 1000 Haar-random realizations of $\mathbf{U}_1$, $\mathbf{U}_2$, and Gaussian samplings of $\mathbf{R}$ in \eqref{eq:A_B_definitions}.} 
\label{fig_capacity_weak_active_transformation}
\end{figure*}

To evaluate the expected capacity one needs to know the distribution of the eigenvalues of $\mathbf{H}_{\textrm{s}}\mathbf{H}_{\textrm{s}}^{\text{T}}$. Since this matrix is a real representation of $\mathbf{A}_1\mathbf{A}_{1}^{\dagger}$, its  eigenspectrum is a doubly degenerate eigenspectrum of the latter matrix, see Appendix~\ref{sec:appendix_SVD_for_optical_transformations}. The matrix $\mathbf{A}_1\mathbf{A}_{1}^{\dagger}$ is Hermitian and has a known joint  probability density of the eigenvalues in a form of the Jacobi ensemble arising from truncated Haar unitary matrix \cite{Forrester2010}: 
\begin{equation}
    \label{eq:Jacobi_basic_ensemble}
    p^{\beta=2}_{\textrm{Jacobi}}(\mathbf{x}) = C_{abK} \prod^{K}_{j=1}x_{j}^{a} (1-x_j)^{b}\prod_{1\leq k <l \leq K} |x_{l}-x_{k}|^{2}
\end{equation}
where $x = (x_1,\dots,x_K) \in [0,1]$, $C_{abK}$ is a normalization factor and $a=N-K$,  $b=M-K$ for $\mathbf{A}_1$ subblock with $K\leq N$. 

In order to derive the capacity one can compute the average spectral density as a marginal of the Jacobi joint probability density: 
\begin{equation}
    p_{\lambda}(\lambda) = \int p^{\beta=2}_{\textrm{Jacobi}}(\lambda,\lambda_2, \dots, \lambda_K) \textrm{d}\lambda_{2} \dots \textrm{d}\lambda_{K}
\end{equation}
which gives \cite{Forrester2010, Livan2011}: 
\begin{equation}
    p_{\lambda}(\lambda) = \frac{\lambda^{a}(1-\lambda)^{b}}{K} \sum^{K-1}_{k=0} \frac{1}{h^{(a,b)}_{k}}\left[\mathcal{P}^{(a,b)}_{k}(1-2\lambda)\right]^{2}
\end{equation}
where $\lambda \in [0,1]$, $\mathcal{P}^{(a,b)}_{k}(1-2\lambda)$ are Jacobi polynomials, and $h^{(a,b)}_{k}$ is the normalization factor: 
\begin{equation}
\label{eq:h_k_definition}
    h^{(a,b)}_{k} = \frac{1}{2k+a+b+1} \frac{(k+a)!(k+b)!}{(k+a+b)!k!}
\end{equation}
The formula for expected information capacity of a random passive Gaussian channel with $N$ transmission modes, $K \leq N$ receiver modes and $M$ environmental modes is therefore given by:
\begin{equation}
    \label{eq:expected_random_capacity_general}
    C = \int_0^{1} C(\lambda)\lambda^{a}(1-\lambda)^{b}\sum^{K-1}_{k=0} \frac{1}{h^{(a,b)}_{k}}\left[\mathcal{P}^{(a,b)}_{k}(1-2\lambda)\right]^{2} \text{d}\lambda,
\end{equation}
where $a=N-K$, $b=M-K$ and $h^{(a,b)}_k$ given by Eq.~\eqref{eq:h_k_definition}. The details and the case $K \geq N$ is presented in Appendix~\ref{sec:appendix_jpd}.

Importantly, note that any Gaussian channel of dimension $N$ can always be represented by a symplectic transformation acting on an additional $M=N$ environmental modes \cite{Caruso2008}. The formula in \eqref{eq:expected_random_capacity_general} simplifies significantly for such a symmetric case $N=K=M$, yielding the capacity
\begin{equation}
\label{eq:expected_random_capacity_symmetric}
   C= \int_{0}^{1} C(\lambda)\sum^{K-1}_{k=0} (2k+1)\left[\mathcal{P}_{k}(1-2\lambda)\right]^{2}\text{d}\lambda,
\end{equation}
where $\mathcal{P}_{k}(x)$ are Legendre polynomials $\mathcal{P}_{k}(x)=\mathcal{P}^{(0,0)}_{k}(x)$.

It is seen in Fig.~\ref{fig_expected_capacities_RMT} that the capacity of a random channel increases with the number of signal modes, irrespectively of the size of the environment. On the other hand, the larger the environment, the lower the capacity. This can be intuitively understood from the fact that random transformations on a large signal+environment system predominantly result in most of the signal lost to the latter subsystem. Note that in the case of homodyne detection, the capacity for the random channel is lower than for heterodyne measurement. This is  because the transmitter has to modulate both quadratures since they do not posses channel state information and therefore cannot recover the singular mode quadrature in which it would be optimal to modulate the signal, see Appendix~\ref{sec:appendix_hetero_homodyne_measurements}.
It is also seen that in the case of no additive noise, the capacity of a random channel is unbounded whereas in the noisy case it saturates at some constant value, depending on the noise strength.

\subsection{Active optical transformations}
\label{sec:active}

When active optical elements are included, the transformation matrices $\mathbf{A},\mathbf{B}$ retain the general form given in Eq.~\eqref{eq:A_B_definitions} with $\mathbf{R}\neq 0$. To model a general random transformation that still defines a valid quantum Gaussian channel, we sample the squeezing parameters $\mathbf{R}=\mathrm{diag}(r_1,\dots,r_{N+M})$ from a Gaussian distribution $\mathcal{N}(0,\sigma^{2})$, while the matrices $\mathbf{U}_1$ and $\mathbf{U}_2$ are drawn independently from the Haar measure. 

Restriction to small $\sigma^{2}$ ensures that the active contribution remains a perturbation of the otherwise passive transformation and that only a small amount of energy is added, this addition is not included in the power constraint for the input signal modulation. 
After sampling the whole transformation and constructing $\mathbf{\Tilde{H}}$ the signal transformation part $\mathbf{H_s}$ is extracted. Then the effective noise matrix $\mathbf{Y} = \tfrac{1}{2}|\Omega_{2K} - \mathbf{H}_{\textrm{s}}\Omega_{2N}\mathbf{H}_{\textrm{s}}^{\text{T}}| + \xi\mathds{1}_{2K}$ is evaluated. The expected value of the information capacity for such model is evaluated using the Monte Carlo simulation method. For each realization of the matrix $\tilde{\textbf{H}}$ generated according to Eq.~\eqref{eq:tilde_H_definition_A_B1}, the channel information capacity is computed from the corresponding output covariance matrix with input power divided equally between available modes. The final expected information capacity is obtained by averaging over many such realizations and is not optimized over input encodings. It is seen in Fig.~\ref{fig_capacity_weak_active_transformation}, that active random transformations decrease the Holevo information relative to the passive case. This is opposite to the heterodyne and homodyne detection for which the capacity is increased by active transformations. It can be seen that for the case with single transmitter and multiple receiver modes, $N=1,K=M$ both Holevo and heterodyne capacities for the random channel attain their noiseless values with growing $K$. Interestingly, in this scenario the homodyne capacity  of the random channel, both passive and active, exceeds the corresponding standard capacity for the identity channel  and approaches the ideal heterodyne capacity as $K$ increases. This is because random mixing distributes information encoded in both quadratures of the single input mode among the measured output quadratures, allowing the receiver to recover information about both quadratures through joint processing of the homodyne outcomes. Additionally, with growing $K$ the fraction of signal that can be recovered after mixing increases. Note that a similar idea cannot be used to increase the homodyne capacity of the ideal channel as for $N=1$ one has only a single occupied output mode.

\section{Conclusions}
\label{sec:Conclusions}

In conclusion, we derived a quantum mechanical version of the capacity of multiple-input multiple-output optical channels. We obtained an exact expression for passive and phase-insensitive channels, which model many practical scenarios. Importantly, the phase sensitive case, which include effects such as one and multimode squeezing is much more involved as the optimal input ensemble crucially depends on the input power, even in the single mode instance \cite{Schafer2016}. 
We showed that passive and phase-insensitive channels may be decomposed into the parallel single-mode channels by an appropriate choice of basis for information encoding. Using such formalism the Holevo information of the Gaussian channel may be maximized using an analogue of the water-filling method. 
We derived the analytical formula for the expected information capacity if the channel is represented by a random passive optical transformation and shown the scaling with the number of modes used for signal encoding. 

\section*{Acknowledgment}
The authors would like to thank Ian Korobov and Konrad Banaszek for insightful discussions.  This work was supported by the project “Quantum Optical Technologies” (contract no. FENG.02.01-IP.05-0017/23) carried out within the Measure 2.1 International Research Agendas programme of the Foundation for Polish Science cofinanced by the European Union under the European Funds for Smart Economy 2021-2027 (FENG). 
 
\appendices

\section{SVD of the Signal Transformation Matrix}
\label{sec:appendix_SVD_for_optical_transformations}

Eq.~\eqref{eq:H_s_phase_insensitive_passive} can be treated as a definition of a real representation of the complex matrix $\mathbf{C} = \mathbf{H}_1 + i\mathbf{H}_2$. We denote this relation as $\mathbf{H}_{\textrm{s}} = \mathcal{R}(\mathbf{C})$, and we use this notation throughout the text. No additional assumptions are made about the structure of $\mathbf{C}$.

One can decompose the complex matrix $\mathbf{C}$ using the SVD:
\begin{equation}
    \mathbf{C} = \mathbf{U}_{\textrm{C}}\mathbf{D}_{\textrm{C}}\mathbf{W}_{\textrm{C}}^{\dagger},
\end{equation}
where $\mathbf{U}_{\textrm{C}}$ and $\mathbf{W}_{\textrm{C}}$ are unitary matrices, and $\mathbf{D}_{\textrm{C}}$ is a real diagonal matrix containing the singular values of $\mathbf{C}$.
By separating the real and imaginary parts of the matrices in the above decomposition, it can be shown that the real representation $\mathcal{R}(\mathbf{C})$ admits the factorization
\begin{equation}
    \mathcal{R}(\mathbf{C}) = \mathcal{R}(\mathbf{U}_{\textrm{C}})\,\mathcal{R}(\mathbf{D}_{\textrm{C}})\,\mathcal{R}(\mathbf{W}_{\textrm{C}})^{\text{T}} = \mathbf{U}\mathbf{D}\mathbf{W}^{\text{T}},
\end{equation}
where
\begin{equation}
    \mathbf{U} = \mathcal{R}(\mathbf{U}_{\textrm{C}}), \quad 
    \mathbf{D} = \begin{bmatrix}
        \mathbf{D}_{\textrm{C}} & 0 \\
        0 & \mathbf{D}_{\textrm{C}}
    \end{bmatrix}, \quad 
    \mathbf{W} = \mathcal{R}(\mathbf{W}_{\textrm{C}}).
\end{equation}
The matrices $\mathbf{U}$ and $\mathbf{W}$ are real representations of the unitary matrices $\mathbf{U}_{\textrm{C}}$ and $\mathbf{W}_{\textrm{C}}$, and are therefore orthogonal and symplectic due to the isomorphism between those groups \cite{Serafini2023}. The matrix $\mathbf{D}$ is block-diagonal, with identical diagonal blocks containing the singular values of $\mathbf{C}$. Since these singular values are real by definition, we have $ \textrm{Re}{\mathbf{D}_{\textrm{C}}}= \mathbf{D}_{\textrm{C}}$ and thus $\mathcal{R}(\mathbf{D}_{\textrm{C}}) = \textrm{Re}{\mathbf{D}_{\textrm{C}}} \oplus \textrm{Re}{\mathbf{D}_{\textrm{C}}} = \mathbf{D}_{\textrm{C}} \oplus \mathbf{D}_{\textrm{C}}$.

Consequently, due to the definition $\mathbf{H}_{\textrm{s}} = \mathcal{R}(\mathbf{C})$, the matrices appearing in the decomposition $\mathbf{H}_{\textrm{s}} = \mathbf{U}\mathbf{D}\mathbf{W}^{\text{T}}$ are orthogonal and symplectic. For rectangular matrices $\mathbf{H}_{\textrm{s}}$, matrix $\mathbf{D}$ is not strictly diagonal along its main diagonal; rather, it is block-diagonal, with each rectangular block being diagonal in the conventional sense. We further refer to this decomposition as the SVD despite this discrepancy for uneven dimensions.

\section{Capacities for Heterodyne and Homodyne Measurements}
\label{sec:appendix_hetero_homodyne_measurements}

For a multimode Gaussian channel in basis corresponding to the canonical form \eqref{eq:diagonal}, the total capacity corresponding to a chosen measurement technique can be decomposed into a sum of capacities of parallel single-mode channels. For the $k$-th mode, one can define the average number of signal photons $\bar{n}_s^{(k)}$used for communication and the mean number of effective noise photons $\bar{n}_n^{(k)}$, respectively, as
\begin{equation}
      \bar{n}_{s}^{(k)} = \lambda_k P_{k}, \quad{}  \bar{n}_n^{(k)} = \frac{\lambda_k - 1}{2} + \left(n + \frac{1}{2}\right)\lvert 1 - \lambda_k \rvert + \xi,
\end{equation}
where $P_k$ denotes the fraction of the total input power $P$ allocated to the $k$-th mode, $\lambda_k$ characterizes the channel transmissivity, $n$ is the environmental thermal photon number, and $\xi$ represents additive noise.

The capacities for single-mode heterodyne and homodyne measurements are then given, respectively, by \cite{Banaszek2020}:
\begin{subequations}
\begin{align}
    C_{k}^{\mathrm{het}} &= \log_2\left(1 + \frac{\bar{n}_s^{(k)}}{\bar{n}_n^{(k)} + 1}\right), \\
    C_{k}^{\mathrm{hom}} &= \frac{1}{2}\log_2\left(1 + \frac{2\bar{n}_s^{(k)}}{\bar{n}_n^{(k)} + \tfrac{1}{2}}\right). \label{eq:appendix_single_mode_hom}
\end{align}
\end{subequations}
This yields the total information capacity of $K$ modes as
\begin{equation}
\label{eq:appendix_het_hom_sum}
    C^{\mathrm{het}} = \sum_{k=1}^{K} C^{\mathrm{het}}_{k}, 
    \quad 
    C^{\mathrm{hom}} = \sum_{k=1}^{K} C^{\mathrm{hom}}_{k},
\end{equation}
which can be further optimized using water-filling technique.

In the simplest case of identity channel, $\forall_k \lambda_k = 1$, with equal power distribution among the input modes, $ \forall_k P_{k} = \tfrac{P}{N}$, an equal number of receiver and transmitter modes, $K = N$, the environment in the vacuum state $n = 0$, and no additive noise $\xi = 0$, the asymptotic behaviour of the capacities as the number of modes $N \to \infty$ is equal to
\begin{subequations}
\begin{align}
    \lim_{N \to \infty} C^{\mathrm{het}} 
    &= \lim_{N \to \infty} N \log_2\left(1 + \frac{P}{N}\right) 
    = \frac{P}{\ln 2}, \\
    \lim_{N \to \infty} C^{\mathrm{hom}} 
    &= \lim_{N \to \infty} \frac{N}{2} \log_2\left(1 + \frac{4P}{N}\right) 
    = \frac{2P}{\ln 2}.
\end{align}
\end{subequations}
Notably, the same limit computed for the Holevo information does not yield a finite bound on the information capacity \cite{Guha2011}.

If the Gaussian channel cannot be decomposed into a set of parallel single-mode subchannels, the information capacities of heterodyne and homodyne measurements must be evaluated using the general mutual information formula for multivariate Gaussian distributions \cite{Pilyavets2008}
\begin{subequations}
\begin{align}
    C^{\textrm{het}} &= \frac{1}{2}\log_2 \det\left[\mathds{1}_{2N}+\mathbf{S}_{\mathrm{het}}\left(\mathbf{N}_{\mathrm{het}}+\frac{1}{2}\mathds{1}_{2N}\right)^{-1}\right],\\
     C^{\textrm{hom}} &= \frac{1}{2}\log_2 \det\left[\mathds{1}_{N}+\mathbf{S}_{\mathrm{hom}}(\mathbf{N}_{\mathrm{hom}})^{-1}\right]
\end{align}
\end{subequations}
where the signal and noise covariances are defined by
\begin{align}
    \mathbf{S}_{\mathrm{het}} &= \mathbf{H}_{\textrm{s}}\mathbf{V}_{\textrm{mod}}\mathbf{H}_{\textrm{s}}^{\textrm{T}}, & 
    \mathbf{N}_{\mathrm{het}} &= \mathbf{H}_{\textrm{s}}\mathbf{H}_{\textrm{s}}^{\textrm{T}}/2+\mathbf{Y},\\
    \mathbf{S}_{\mathrm{hom}} &= \mathbf{\Pi} \mathbf{S}_{\textrm{het}} \mathbf{\Pi}^{\mathrm{T}}, & 
    \mathbf{N}_{\mathrm{hom}} &= \mathbf{\Pi} \mathbf{N}_{\textrm{het}} \mathbf{\Pi}^{\mathrm{T}},
\end{align}
where $\mathbf{V}_{\textrm{mod}}$ denotes covariance matrix of the signal modulation and $\mathbf{\Pi}$ is the projection onto the measured quadratures, e.g., $\mathbf{\Pi} = [\mathds{1}_N, 0_N]$ for homodyne measurement of $\hat{q}$.

For channels with diagonal covariance matrices, diagonalized $\mathbf{H}_{\textrm{s}}$ and equal power distribution between quadratures per mode, the heterodyne capacity reduces to the sum of single-mode channel capacities in Eq.~\eqref{eq:appendix_het_hom_sum}. Similarly, if the modulation power is concentrated only in a single measured quadrature type, the homodyne capacity is recovered from the same equation. Note, that in the case of homodyne detection, the modulated quadratures have to correspond to normal modes of the channel transmission matrix $\mathbf{H}_{\textrm{s}}$ in order to obtain the highest capacity, otherwise, due to mixing between different modes and quadratures the capacity will be different. However, if the power is distributed uniformly between the quadratures, as is the case in the Sec.~\ref{sec:random}, in the basis that is being measured, the single mode formula for homodyne measurement capacity Eq.~(\ref{eq:appendix_single_mode_hom}) should not have a factor $2$ in the signal-to-noise ratio in the numerator. This caveat is not present in case of heterodyne measurement since then both quadratures are always measured.

\section{Joint Probability Distribution of $\mathbf{A}_{1}\mathbf{A}_{1}^{\dagger}$}
\label{sec:appendix_jpd}

According to Proposition 3.8.3 in \cite{Forrester2010}, let $\mathbf{A}_0$ be an $n_1 \times n_2$ 
($n_1 \geq n_2$) subblock of a Haar-distributed unitary matrix $\mathbf{U} \in U(N+M)$. Then the joint probability density of the eigenvalues of the matrix $\mathbf{A}_0^{\dagger}\mathbf{A}_0$ is given by the Jacobi ensemble
\begin{equation}
    \label{eq:Jacobi_basic_ensemble_appendix}
    p^{\beta=2}_{\mathrm{Jacobi}}(\mathbf{x})
    = C_{a b \beta}
    \prod_{j=1}^{n_2} x_j^{a}
    (1-x_j)^{b}
    \prod_{1\le k<l\le n_2} |x_l - x_k|^{2},
\end{equation}
where $x = (x_1,\dots,x_K) \in [0,1]$, $a = n_1 - n_2$, $b = (N+M) - n_1 - n_2$.

We now consider the first case relevant to our setting. Let $\mathbf{A}_1$ denote an $n_2 \times n_1$ ($n_1 \geq n_2$) subblock of a Haar-distributed unitary matrix in $U(N+M)$. We are interested in the joint probability density of the eigenvalues of $\mathbf{A}_1\mathbf{A}_1^{\dagger}$. Since the Haar measure on $U(N+M)$ is invariant under taking adjoints, the matrix $\mathbf{A}_1^{\dagger}$ is an $n_1 \times n_2$ subblock of a Haar-distributed unitary matrix of the same dimension. Defining $\mathbf{A}_0 := \mathbf{A}_1^{\dagger}$, we observe that $\mathbf{A}_0^{\dagger}\mathbf{A}_0=\mathbf{A}_1\mathbf{A}_1^{\dagger}$. Therefore, the distribution of $\mathbf{A}_1\mathbf{A}_1^{\dagger}$ coincides with the Jacobi ensemble given above, with the same parameters $a,b$. Hence, for $\dim \mathbf{A}_1 = K \times N$ with $N \geq K$, we identify that the joint probability distribution of $\mathbf{A_1}\mathbf{A_1}^{\dagger}$ is given by Eq.~\eqref{eq:Jacobi_basic_ensemble_appendix} with $n_1 = N$, $n_2 = K$, which gives $a=N-K$,  $b=M-K$.

We now consider the second case, when $\mathbf{A}_1$ has dimensions $K\times N$ with $K \geq N$. The joint probability density of eigenvalues of $\mathbf{A}_1^{\dagger}\mathbf{A}_1$ is given by Eq.~\eqref{eq:Jacobi_basic_ensemble_appendix}. We are interested in the joint probability distribution of $\mathbf{A}_1\mathbf{A}_1^{\dagger}$, which is a $K \times K$ matrix. Since $\mathrm{rank}(\mathbf{A}_1) \leq N$, this matrix has exactly $N$ eigenvalues, which coincide with those of $\mathbf{A}_1^{\dagger}\mathbf{A}_1$, while the remaining $K-N$ eigenvalues are equal to zero. This allows one to rewrite Eq.~\eqref{eq:expected_value_capacity_RMT}:
\begin{multline}
    \sum_{j=1}^{K}\mathds{E}\left[C(\lambda_j)\right] = \sum_{j=1}^{N}\mathds{E}\left[C(\lambda_j)\right] +  \sum_{j=N+1}^{K}\mathds{E}\left[C(\lambda_j)\right] = \\ = N \int C(\lambda) p_{\lambda}(\lambda)\textrm{d}\lambda + \sum_{j=N+1}^{K}\mathds{E}\left[C(0)\right] =  N \int C(\lambda) p_{\lambda}(\lambda)\textrm{d}\lambda
\end{multline}
where $p_{\lambda}(\lambda)$ is the marginal of the joint probability density of the form given by Eq.~\eqref{eq:Jacobi_basic_ensemble_appendix} with $a=K-N$, $b=M-K$. The last equality follows from the fact that $C(0)$ corresponds to the information capacity of a single mode channel with zero transmission and therefore vanishes.

The expected information capacity formula accounting for both cases can be written as:
\begin{equation}
 C= \sum_{j=1}^{K}\mathds{E}\left[C(\lambda_j)\right] = \min(K,N) \int_{0}^{1} C(\lambda) p_{\lambda}(\lambda)\textrm{d}\lambda,
\end{equation}
where $p_{\lambda}(\lambda)$ is an average spectral density obtained by calculating the marginal of the joint probability distribution of the form of Eq.~\eqref{eq:Jacobi_basic_ensemble_appendix} with $n_2 = \min(K,N), n_1 = \max(K,N)$:
\begin{equation}
    \label{eq:spectral_Jacobi_density_appendix}
        p_{\lambda}(\lambda) = \frac{\lambda^{a}(1-\lambda)^{b}}{\min(K,N)} \sum^{\min(K,N)-1}_{k=0} \frac{1}{h^{(a,b)}_{k}}\left[\mathcal{P}^{(a,b)}_{k}(1-2\lambda)\right]^{2}
\end{equation}
with $a=\max(K,N)-\min(K,N),b=N+M-\min(K,N)-\max(K,N)$.

\bibliographystyle{IEEEtran}
\bibliography{IEEEabrv,Bibliography}
\end{document}